\documentclass[a4paper,10pt,twoside]{cpc-hepnp}

\usepackage{multicol}
\usepackage{graphicx}
\usepackage{booktabs}
\usepackage{amssymb,bm,mathrsfs,bbm,amscd}
\usepackage[tbtags]{amsmath}
\usepackage{lastpage}

\begin{document}

\fancyhead[c]{\small Submitted to `Chinese Physics C'} \fancyfoot[C]{\small 010201-\thepage}

%\footnotetext[0]{Received 14 March 2015}

\title{The effects of $\delta$ mesons on the baryonic direct Urca processes in neutron star matter\thanks{Supported by Natural Science
Foundation of China under grants Nos.11447165 }}

\author{%
HUANG~Xiu-Lin$^{1;1)}$%
\quad WANG~Hai-Jun$^{1)}$
\quad LIU~Guang-Zhou$^{1}$
\quad XU Yan$^{2;2)}$\email{xuy@cho.ac.cn}
}
\maketitle

\address{%
$^1$  Center for Theoretical Physics, Jilin University, Changchun 130012, China\\
$^2$  Changchun Observatory, National Astronomical Observatories,
Chinese Academy of Sciences, Changchun 130117, China}

\begin{abstract}
In the framework of relativistic mean field theory, the relativistic
neutrino emissivity of the nucleonic and hyperonic direct Urca
processes in the degenerate baryon matter of neutron stars are
studied. We investigate particularly the influence of the isovector
scalar interaction which is considered by exchanging $\delta$ meson
on the nucleonic and hyperonic direct Urca processes. The results
indicate that $\delta$ mesons lead to obvious enhancement of the
total neutrino emissivity, which must result in more rapid cooling
rate of neutron star matter.
\end{abstract}



\footnotetext[0]{\hspace*{-3mm}\raisebox{0.3ex}{$\scriptstyle\copyright$}

}%

\begin{multicols}{2}

\section{Introduction}

Compact stellar object studies allow us to build a better and more
comprehensive understanding of the properties of nuclear matter
under extreme conditions. Neutron star (NS) is an ideal model for
study of dense matter physics. It is the remnants of a supernova
explosion with an internal temperature as high as $10^{11}$
¨C$10^{12}$ K, and cooling gradually to $10^{10}$ K by emitting
neutrinos within minutes\cite{lab1,lab2,lab3,lab4,lab5}. Then the
long-term cooling process is mainly through neutrino emission in NS
interior for about $10^{6}$ years. The primary mechanisms in the
cooling stage can be divided into two categories: enhanced neutrino
processes, which include the nucleonic direct Urca (NDURCA)
processes and hyperonic direct Urca (YDURCA) processes. And standard
neutrino processes (such as modified Urca processes and
Bremsstrahlung processes). NS cooling properties tend to be
dominated by whichever reaction has the highest neutrino emissivity.
It is well known that NDURCA processes produce the most powerful
neutrino emissivity, the second is YDURCA processes in the NS core,
which are more efficiently than standard
processes\cite{lab6,lab7,lab8}. While in these work, the formulae of
neutrino emissivity for NDURCA and YDURCA processes are still in
non-relativistic manner. In fact, the threshold densities for NDURCA and YDURCA
processes are larger than the nuclear saturation
density. Above the threshold densities, the baryonic motion is
relativistic in NS core. Thus the formulae in relativistic manner
should be use to calculate the neutrino emissivity for NDURCA and
YDURCA processes. Meanwhile the equation of state (EOS) must be
relativistic, which can be consistent with the relativistic
approach. The relativistic mean field theory (RMFT) is an effective
model for studying the properties of NS\cite{lab9,lab10,lab11,lab12,lab13,lab14,lab15}. The standard RMFT
includes isoscalar-scalar meson $\sigma$, isoscalar-vector meson
$\omega$ and isovector-vector meson $\rho$. The isovector scalar
channel usually is not contained because it is not expected to be
essential to nuclei. However, the isovector scalar interaction could
have an important role for asymmetric nuclear matter. It can be
introduced through a coupling to $\delta$ ($a_0(980)$) meson which
was studied by many
authors\cite{lab16,lab17,lab18,lab19,lab20,lab21,lab22,lab23,lab24}.
These studies have shown that $\delta$ mesons had definite
contributions to isospin-asymmetric nuclear matter, especially at
high densities. NSs keep away from isospin-symmetric nuclear matter
due to charge neutrality. Therefore, in this work we extend the
analysis the contribution of the $\delta$ field in dense asymmetric
matter to the effect on NS matter. When an NS includes $\delta$
mesons, the $\delta$ field not only changes the abundance of
neutrons, protons and hyperons, but also changes the bulk properties
of NS, which must affect the neutrino emissivity of NDURCA and
YDURCA processes.

In this work, we use the RMFT including $\sigma$, $\omega$, $\rho$
and $\delta$ mesons with additional cubic and quartic nonlinearities
of $\sigma$ meson to describe baryons interacting. The $\sigma$ and
$\omega$ mesons supply medium-range attractive and short-range
repulsive interactions between baryons, respectively. The $\delta$
and $\rho$ mesons provide the corresponding attractive and repulsive
potentials in the isovector channel, respectively. We adopt the
simplest NS model, assuming NS core consists of n, p, $\Lambda$,
$\Sigma^-$, $\Sigma^0$, $\Sigma^+$, $\Xi^-$, $\Xi^0$, e and $\mu$.
This work mainly focuses on the effects of the $\delta$ meson on the
total neutrino emissivity for NDURCA and YDURCA processes in NS
core.

\section{Models}
The effective Lagrangian function for the RMFT can be written as follows:
\begin{equation}\label{1}
\begin{split}
    \mathcal{L}=&\sum_B \overline{\Psi}_B[i \gamma_\mu \partial^\mu-(m_{B}-g_{\delta B} \bm{\tau} \cdot \bm{\delta}-g_{\sigma B}\sigma)-g_{\omega B}\gamma_\mu \omega^\mu \\
    &-g_{\rho B} \bm{\tau} \cdot {\bm{\rho}}^\mu]\Psi_B+\frac{1}{2}(\partial_\mu \sigma \partial^\mu \sigma-m_\sigma^2 \sigma^2)-U(\sigma)\\
    &+\frac{1}{2}m_\omega^2 \omega_\mu \omega^\mu+\frac{1}{2}m_\rho^2 \bm{\rho}_\mu \bm{\rho}^\mu+\frac{1}{2}(\partial_\mu \bm{\delta} \partial^\mu \bm{\delta}-m_\delta^2 \bm{\delta}^2)\\
    &+\frac{1}{4}F_{\mu \nu} F^{\mu \nu}-\frac{1}{4}\bm{G}_{\mu \nu} \bm{G}^{\mu \nu}+\sum_l \overline{\Psi}_l[i \gamma_\mu \partial^\mu-m_{B}]\Psi_l,\\
\end{split}
\end{equation}
where the potential function
$U(\sigma)=\frac{1}{3}a\sigma^3+\frac{1}{4}b\sigma^4$, $F_{\mu
\nu}=\partial_\mu \omega_\nu-\partial_\nu \omega_\mu$ and
$\bm{G}_{\mu \nu}=\partial_\mu \bm{\rho}_\nu-\partial_\nu
\bm{\rho}_\mu$.

In the RMFT, the meson fields can be considered as classical
fields, where the field operators are replaced by their expectation
values. Meson field equations are obtained as follows:
\begin{equation}\label{2}
    m_\sigma^2 \sigma+a \sigma^2+b \sigma^3=\sum_B \frac{g_{\sigma B}}{\pi^2} \int_0^{p_B}\frac{m^\ast_B k^2}{\sqrt{k^2+{m^\ast_B}^2}}dk,
\end{equation}
\begin{equation}\label{3}
    m_\omega^2 \omega_0 =\sum_B g_{\omega B} \rho_B,
\end{equation}

\begin{equation}\label{4}
    m_\rho^2 \rho_0 =\sum_B g_{\rho B} I_{3B} \rho_B,
\end{equation}

\begin{equation}\label{5}
    m_\delta^2 \sigma_0=\sum_B \frac{g_{\delta B}}{\pi^2} \int_0^{p_B}\frac{m^\ast_B k^2}{\sqrt{k^2+{m^\ast_B}^2}}dk,
\end{equation}
where $p_B$ is the baryonic Fermi momentum, $I_{3B}$
is the baryonic isospin projection, the baryonic
density is expressed by:
\begin{equation}\label{6}
  \rho_B=\frac{p_B^3}{3\pi^2}.
\end{equation}
The baryonic effective mass $m_B^\ast$ is written:
\begin{equation}\label{6}
    m_B^\ast=m_B-g_{\sigma B}\sigma_0-I_{3B}g_{\sigma B}\delta_0.
\end{equation}

Under $\beta$ equilibrium conditions, the chemical potentials of
baryons and leptons satisfy the following relationship:
\begin{equation}\label{8}
    \mu_B=\mu_n-q_B \mu_e, \mu_\mu=\mu_e.
\end{equation}
At zero temperature approximation, they are expressed as follows:
\begin{equation}\label{8}
    \mu_B=\sqrt{p_B^2+{m_B^\ast}^2}+g_{\omega B} \omega_0+g_{\rho B} \rho_0 I_{3B},
\end{equation}

\begin{equation}\label{9}
    \mu_l=\sqrt{p_l^2+{m_l^\ast}^2}.
\end{equation}
NSs also need to meet the conditions of electrical neutrality and
baryon number conservation, which are expressed as follows:
\begin{equation}\label{7}
    \sum_B q_B \rho_B-\rho_e-\rho_\mu=0,
\end{equation}
\begin{equation}\label{7.1}
    \sum_B \rho_B=\rho,
\end{equation}
where $\rho$ is total baryonic density.

The possible neutrino emission processes consist of two successive
reactions, beta decay and capture, expressed as follows:
\begin{equation}\label{9}
    B_1 \rightarrow {B}_2+l+{\overline{\nu}}_l, B_2+l \rightarrow {B}_1+{\nu}_l,
\end{equation}
where $B_1$ and $B_2$ are baryons, and $l$ is a lepton. This paper
focuses on electron processes. The relativistic neutrino emissivity
can be given by Fermi golden rule. The relativistic expression of
the energy loss $Q_R$ per unit volume and time is expressed as\cite{lab25}:
\begin{equation}\label{10}
\begin{split}
    Q_R=& \frac{457\pi}{10080}G_F^2 C^2 T^6 \Theta (p_e+p_{B_2}-p_{B_1}) \\
    & \times \{ f_1 g_1[(\varepsilon_{F_1}+\varepsilon_{F_2})p_e^2-(\varepsilon_{F_1}-\varepsilon_{F_2})(p^2_{B_1}-p^2_{B_2})]\\
    &+2g_1^2 \mu_e m^\ast_{B_1} m^\ast_{B_2}+(f^2_1+g^2_1)[\mu_e(2\varepsilon_{F_1}\varepsilon_{F_2}-m^\ast_{B_1}m^\ast_{B_2})\\
    &+\varepsilon_{F_1}P_e^2-\frac{1}{2}(p^2_{B_1}-p^2_{B_2}+p^2_e)(\varepsilon_{F_1}+\varepsilon_{F_2})],\\
\end{split}
\end{equation}

\begin{center}
\tabcaption{ \label{tab2}The constants of NDURCA and YDURCA
processes. $sin \theta_c=0.231\pm0.003$, $F=0.477\pm0.012$,
$D=0.756\pm0.011$} \footnotesize
\begin{tabular*}{80mm}{c@{\extracolsep{\fill}}cccc}
\toprule  Process & Transition & $C$   & ${f}_1$  & ${g}_1$ \\
\hline
1\hphantom{0} & \hphantom{0}$n \rightarrow p e {\bar{\nu}}_e$ & cos$\theta_C$ & 1 & F+D\\
2\hphantom{0} & \hphantom{0}$\Lambda \rightarrow p e {\bar{\nu}}_e$ & sin$\theta_C$ & $-\sqrt{3/2}$ & $-\sqrt{3/2}(F+D/3)$\\
3\hphantom{0} & $\Sigma^{-} \rightarrow n e {\bar{\nu}}_e$ & sin$\theta_C$ & -1 & $-(F-D)$\\
4\hphantom{0} & $\Sigma^{-} \rightarrow \Lambda e {\bar{\nu}}_e$ & cos$\theta_C$ & 0 & $\sqrt{3/2}D$\\
5\hphantom{0} & $\Sigma^{-} \rightarrow \Sigma^0 e {\bar{\nu}}_e$ & cos$\theta_C$ & $\sqrt{2}$ & $\sqrt{2}F$\\
6\hphantom{0} & $\Xi^{-} \rightarrow \Lambda e {\bar{\nu}}_e$ & sin$\theta_C$ & $\sqrt{3/2}$ & $\sqrt{3/2}(F-D/3)$\\
7\hphantom{0} & $\Xi^{-} \rightarrow \Sigma^0 e {\bar{\nu}}_e$ & sin$\theta_C$ & $\sqrt{1/2}$ & $(F+D)/\sqrt{2}$\\
8\hphantom{0} & $\Xi^{-} \rightarrow \Sigma^+ e {\bar{\nu}}_e$ & sin$\theta_C$ & 1 & $F+D$\\
9\hphantom{0} & $\Xi^{-} \rightarrow \Xi^0 e {\bar{\nu}}_e$ & cos$\theta_C$ & 1 & $F-D$\\
\bottomrule
\end{tabular*}
\end{center}

\begin{center}
\tabcaption{\label{tab2} Parameter set. $f_i=(g_i /
m_i)^2(fm^2)$,$i=\sigma$, $\omega$, $\rho$, $\delta$, and
$A=a/g_\sigma^3 (fm^{-1})$, $B=b/g_\sigma^4$} \footnotesize
\begin{tabular*}{80mm}{ccccccc}
\toprule Parameter & $f_\sigma $  & $f_\omega$  & $f_\rho$ & $f_\delta$ &A&B \\
\hline
without $\delta$ & 10.33 & 5.42 & 0.95& 0.00& 0.033& -0.0048 \\
with  $\delta$  & 10.33 & 5.42 & 0.95& 2.50& 0.033& -0.0048 \\
\bottomrule
\end{tabular*}
\end{center}

\begin{center}
\includegraphics[width=8.5cm]{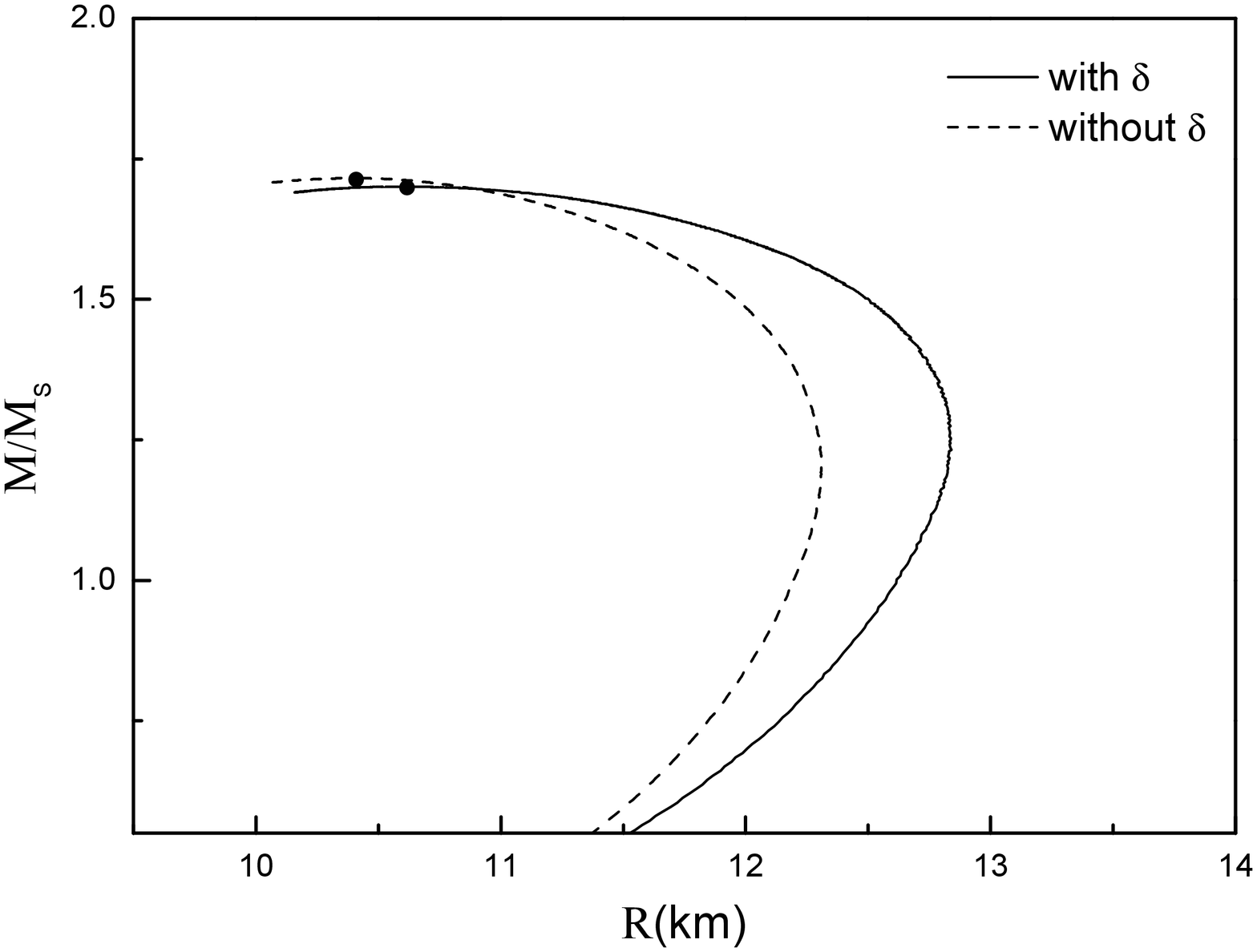}
\figcaption{\label{fig1}The mass-radius relations for NSs with and
without $\delta$ mesons. The dots stand for the NSs maximum masses
for the two cases.}
\end{center}
\begin{center}

\tabcaption{ \label{tab3}  The maximum masses $M_{max}$, the
corresponding radius R and center density $\rho_c$ with and without $\delta$
mesons.} \footnotesize
\begin{tabular*}{80mm}{c@{\extracolsep{\fill}}cccc}
\toprule  Parameter & $M_{max} / M_{s}$  &R & $\rho_c$\\
\hline
 without $\delta$ & 1.72  & 10.40&1.23\\
 with  $\delta$  & 1.70  & 10.58&1.20\\
\bottomrule
\end{tabular*}
\end{center}
where $G_F=1.436 \times 10^{-49}$ erg $cm^3$ is the weak-coupling
constant, $f_1$ and $g_1$ are the vector and axial-vector constants,
and $C$ is the Cabibbo factor. $\mu_e$ is the chemical potential of
an electron, and $p_{B_1}$, and $p_{B_2}$, $p_{e}$ are the
Fermi-momenta of baryons $B_1$, $B_1$, and electrons, respectively.
$\varepsilon_{F_1}$ and $\varepsilon_{F_2}$ are the kinetic energy
of baryons $B_1$ and $B_1$. When $x\geq0$, $\Theta(x)=1$, and when
$x<0$, $\Theta(x)=0$. The parameters $C$, $f_1$, and $g_1$ for
NDURCA and YDURCA in NS matter are listed in Table 1\cite{lab26}.

Solving the coupling equations self-consistently at a fixed baryon
density $\rho$, one can obtain a list of physical quantities,
such as the particle fraction, effective mass, nucleon Fermi
momentum. Then the total relativistic neutrino emissivity for NDURCA
and YDURCA processes can be obtained.

\section{Discussion}

A primary contribution for NSs including $\delta$ mesons is changing
the EOS, which must lead to the change of the bulk properties of
NSs\cite{lab17,lab18,lab19}. We substitute EOS into the
Tolman-Oppenheimer-Volkoff (TOV) equations\cite{lab27,lab28}, the
mass-radius relations for NSs can also be obtained. The most
important physical quantities for the neutrino emissivity of NDURCA
and YDURCA processes are the particle fraction, Fermi momentum and
effective mass of nucleon and hyperon. This article focuses on the
effect of the inclusion of $\delta$ mesons on NDURCA and YDURCA
processes in NSs. Next, we will give the numeric results of the
relativistic neutrino emissivity for NDURCA and YDURCA processes
with and without $\delta$ mesons. The properties of NSs are obtained
with the parameter sets presented in Table 2. The hyperon couplings
are represented as the ratios to the nucleon coupling $g_{\sigma
H}=0.7 g_{\sigma N}$, $g_{\omega H}=0.7 g_{\omega N}$, $g_{\rho
H}=0.7 g_{\rho N}$, and $g_{\delta H}=0.7 g_{\delta N}$.

The mass-radius relations of NSs with and without $\delta$ mesons
are compared in Fig. \ref{fig1}.
The maximum masses $M_{max}$, the
corresponding radius R and center density $\rho_c$ with and without $\delta$ mesons are
displayed in Table 3. From Fig. 1 and Table 3, one can find that the
inclusion of $\delta$ mesons leads to the larger radius for an NS
with the same mass.

\begin{center}
\includegraphics[width=8.5cm]{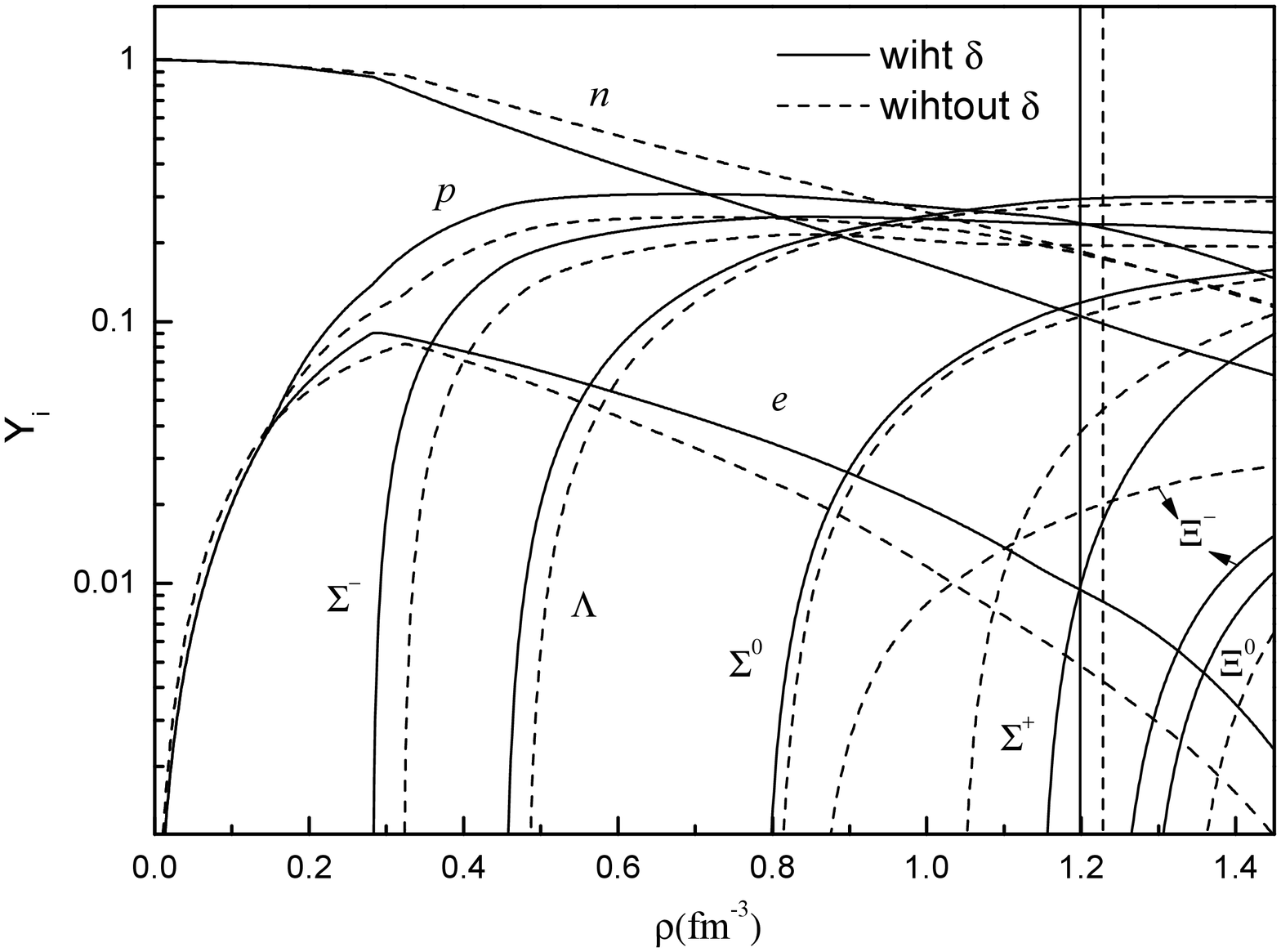}
\figcaption{\label{fig2}The particle fractions of baryons and
leptons as a function of the baryon density $\rho$ with and
without $\delta$ mesons. The two vertical lines represent the
maximum center densities of the maximum masses NS with and without
$\delta$ mesons.}
\end{center}

The particle fraction $Y_i$ as a function of the baryonic density
$\rho$, with or without $\delta$ mesons is shown in Fig. 2. As
seen in Fig. 2, whether or not the $\delta$ meson is included, the
$\Sigma$, $\Lambda$ and $\Xi$ appear one by one. This is because
that a hyperonic type is populated only if its chemical
potential exceeds its lowest energy state in NS matter, e.g.
$\mu_n-q_B\mu_e\geq m_B^\ast+g_{{\omega}B}\omega_0+g_{\rho B}
{\rho}_{0} {I}_{3B}$. So when the thresholds are reached, additional
hyperonic species are populated. As shown in Fig.2, the inclusion of
$\delta$ mesons changes the baryonic threshold conditions, the
threshold densities of p, $\Lambda$, $\Sigma^0$, $\Sigma^-$ and
$\Xi^0$ with $\delta$ mesons are shifted to lower densities. And the fractions $Y_{p}$, $Y_{\Lambda}$,
$Y_{\Sigma^0}$, $Y_{\Sigma^-}$ and $Y_{\Xi^0}$ with $\delta$ mesons
are larger than the corresponding values without $\delta$ mesons.
Then according to Eq. (6), when the $\delta$ field is included,
$p_{p}$, $p_{\Lambda}$, $p_{\Sigma^0}$, $p_{\Sigma^-}$ and
$p_{\Xi^0}$ are larger than the corresponding values without the
$\delta$ field. From Fig. 2, we can also see that the inclusion of $\delta$ mesons makes the
fractions $Y_{n}$, $Y_{\Sigma^+}$ and $Y_{\Xi^-}$ decrease due to
the charge neutrality and $\beta$-equilibrium conditions. The
changes of baryonic fractions must change the neutrino emissivity
of NDURCA and YDURCA processes.

\begin{center}
\includegraphics[width=8.5cm]{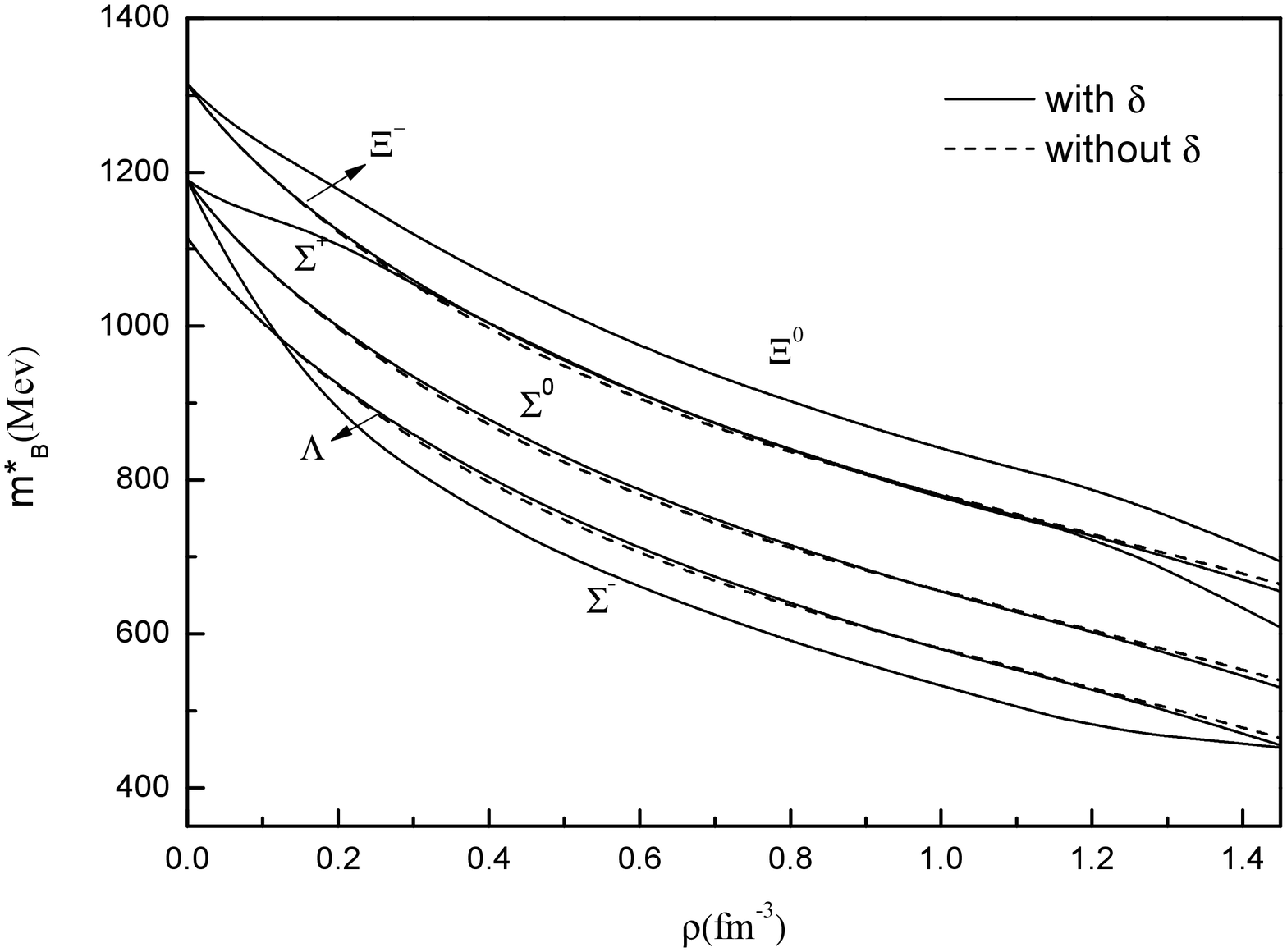}
\figcaption{\label{fig3}The hyperonic effective masses as a function
of the baryonic density $\rho$ with and without $\delta$
mesons.}
\end{center}

Fig. 3 represents the hyperonic effective masses of hyperons as a
function of the baryonic density $\rho$ with and without $\delta$
mesons. According to Eq. (7), the $\delta$ field strength is directly
related to the baryonic effective mass. From Fig. 3, one can find
that the effective mass splitting for hyperons with similar species
but different isospins. The relativistic neutrino emissivity of
NDURCA and YDURCA processes as a function of the baryonic density
$\rho$ with and without $\delta$ mesons are plotted in Fig. 4. One
can find in Fig. 4 the different direct Urca reactions 1-9 in Table 1
appear with increasing of the baryonic density $\rho$. When the
reaction 4 (or 5) happens in NS matter with (or without) $\delta$
mesons, the relativistic neutrino emissivity $Q_{R}$ reaches a
maximum value. While reactions 6-9 (or 8 and 9) with (or without)
$\delta$ mesons would never have happened within stable NSs, because
they occur at higher densities which is larger than the center
densities of the maximum masses NS. As shown in Fig. 4, whether or
not the $\delta$ meson is included in NS matter, reactions 2 and 3
occur as long as $\Lambda$ and $\Sigma^-$ hyperon appear. Namely,
the triangle condition $p_{B_2}+p_{E}>p_{B_1}$ in Eq. (14) is
satisfied automatically for reactions 2 and 3 if $\Lambda$,
$\Xi^{-}$ hyperons appear in NS core. The occurrence of reactions 2
and 3 does not need $Y_\Lambda$, $Y_\Sigma^{-}$ reach a certain
quantity, they are much more likely to happen than the other direct
Urca reactions. While the reactions 4 and 5 (or 4-7) with (or without)
$\delta$ mesons occur later.  The reactions 6 and 7 only occur in
NSs without inclusion of $\delta$ mesons. As shown in Fig. 4, when
$\delta$ mesons are included, the relativistic neutrino emissivity
$Q_{R}$ is obviously larger than the corresponding values without
$\delta$ mesons. It may be concluded that the inclusion of $\delta$
mesons would accelerate the nonsuperfluid NS cooling in most mass
ranges of happening NDURCA and YDURCA processes.

\begin{center}
\includegraphics[width=8.5cm]{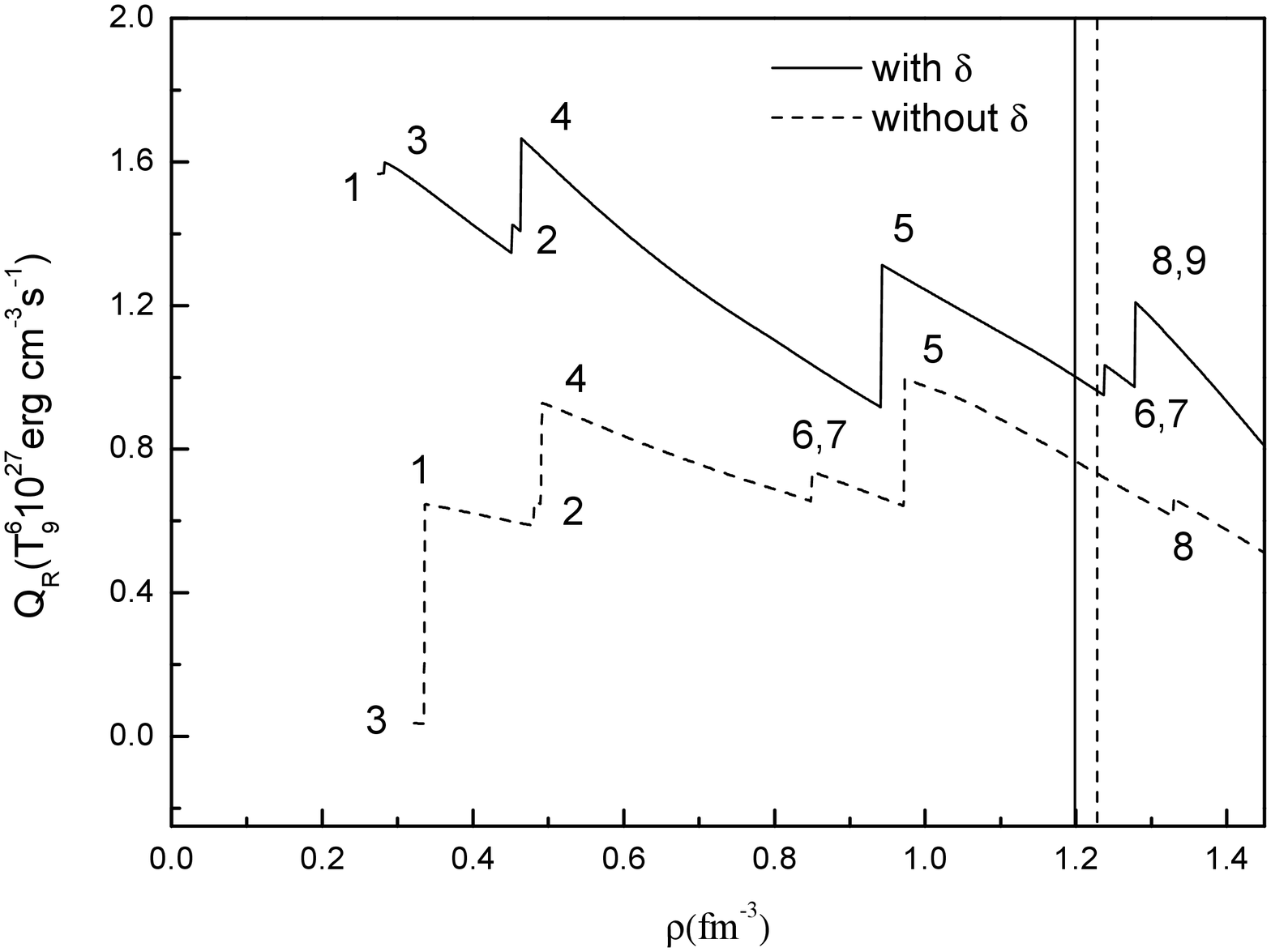}
\figcaption{\label{fig4}The relativistic neutrino emissivity $Q_{R}$
as a function of the baryonic density $\rho$ with and without
$\delta$ mesons in NS matter. The vertical two lines represent the
center densities of the maximum masses NS with and without $\delta$
mesons.}
\end{center}

\section{Conclusions}

We have shown the influence of $\delta$ mesons on the relativistic
neutrino emissivity for NDURCA and YDURCA processes by adopting the
RMFT in NS matter. Results show that the inclusion of $\delta$
mesons makes the baryonic threshold densities change in NS matter,
which lead to the threshold densities of reactions 1-5 changing.
Furthermore, the inclusion of $\delta$ mesons makes the fractions
6-9 with $\delta$ mesons would never have occurred within stable NS
matter. The fractions of p, $\Lambda$, $\Sigma^0$, $\Sigma^-$
increase, which leads to an obvious enhancement of the relativistic
neutrino emissivity in NS matter. Thus the $\delta$ mesons would
speed up the nonsuperfluid NS cooling rate in most mass ranges of
happening NDURCA and YDURCA processes.

\vspace{-1mm}
\centerline{\rule{80mm}{0.1pt}}

\end{multicols}

\clearpage

\end{document}